%% file: 11Geneprofilingpaper.tex
\documentclass[11pt]{article}

\usepackage{graphicx} 
\usepackage{amsmath, amsthm, amsfonts, amscd,}
\usepackage{url}
\usepackage{verbatim}
\usepackage{natbib}
\usepackage{color}
\usepackage{fp}
\usepackage[textwidth=146mm,textheight=215mm]{geometry}
\usepackage{pstricks}

\graphicspath{{./Figures/}}

\numberwithin{equation}{section}

\begin{document}
\title{Gene profiling for determining pluripotent genes in a time course microarray experiment}
\author{J. TUKE\footnote{To whom correspondence should be addressed}\\
{\it School of Mathematical Sciences}\\ 
{\it The University of Adelaide, Adelaide SA 5005, Australia}\\ simon.tuke@adelaide.edu.au\\\\
G.F.V. GLONEK\\ 
{\it School of Mathematical Sciences}\\
{\it The University of Adelaide, Adelaide SA 5005, Australia}\\ gary.glonek@adelaide.edu.au\\\\ 
P.J. SOLOMON\\
{\it School of Mathematical Sciences}\\
{\it The University of Adelaide, Adelaide SA 5005, Australia}\\ patty.solomon@adelaide.edu.au}
\date{}
\maketitle

\section{Abstract}
In microarray experiments, it is often of interest to identify genes which have a pre-specified gene expression profile with respect to time. Methods available in the literature are, however, typically not stringent enough in identifying such genes, particularly when the profile requires equivalence of gene expression levels at certain time points. In this paper, the authors introduce a new methodology, called gene profiling, that uses simultaneous differential and equivalent gene expression level testing to rank genes according to a pre-specified gene expression profile. Gene profiling treats the vector of true gene expression levels as a linear combination of appropriate vectors, i.e., vectors that give the required criteria for the profile. This gene-profile model is fitted to the data and the resultant parameter estimates are summarized in a single test statistic that is then used to rank the genes. The theoretical underpinnings of gene profiling (equivalence testing, intersection-union tests) are discussed in this paper, and the gene profiling methodology is applied to our motivating stem cell experiment. 
\\\\
\noindent{\it Keywords}: Gene expression; Gene profiling; Linear model; Microarray; Pluripotency; Stem cell; Time course experiment.


\section{Introduction}
\label{sec:introduction}
Microarray technology enables researchers to examine the expression levels for many thousands of genes simultaneously \citep[see, for example,][]{Nguyen:2002,Smyth:2003}. Increasingly, information on gene expression is used to infer cell protein levels and thus cellular behaviour \citep{Nguyen:2002,Smyth:2003,Ahnert:2006,McLachlan:2006}. A further major area of interest is in investigating changes in gene expression levels over time in a population of cells \citep{Dudoit:2002,Bar-Joseph:2003,Glonek:2004,Tai:2005,Ernst:2005,Brown:2006,Ahnert:2006} and this is the subject of the present paper. We refer to the gene expression levels over time as a gene expression profile, or profile for short.  
\par
Several methods of analysing gene expression profiles fall into the class of techniques known as unsupervised learning methods. These methods seek to group genes into a number of classes based upon their observed profiles. 
Some of the methodologies discussed in the recent microarray literature are hierarchical classification \citep{Eisen:1998}, self-organizing maps \citep{Tamayo:1999}, the $K$-means algorithm \citep{Tavazoie:1999}, multivariate Gaussian mixtures \citep{Ghosh:2002,Yeung:2001}, and mixtures of linear mixed models \citep{Celeux:2005}.
A related problem that arises in applications of microarray time course experiments is to specify, in advance, a gene expression profile of interest and then to identify the genes with matching expression profiles. However, unsupervised methods do not address this problem and various alternative approaches have been proposed. 
\par
One such method is Pareto optimization, proposed by  \cite{Fleury:2002} and \cite{Hero:2004}, in which a set of functions, each measuring the association of a gene to a pre-specified profile, is chosen. Genes found to be Pareto-optimal with respect to these criteria are identified as matching the pre-specified profile. The main disadvantage with Pareto optimization is that some genes will be selected as Pareto-optimal genes whilst only matching the pre-specified profile for a subset of the profile's criteria.
\par
In an unpublished paper, \cite{Lonnstedt:2006} describe a different method for ranking genes, based on the inner product between the vector of observed log ratios and a pre-specified profile. This method works well for some profiles, but did not provide useful outcomes in our application. 
\par
Gene profiling is a new approach developed by the present authors, which aims to identify genes that match a pre-specified gene expression profile, with greater specificity than the previously described approaches. Gene profiling entails treating the vector of true gene expression levels for each gene as a linear combination of linearly independent vectors chosen to represent the pre-specified profile. The gene-profile model is fitted to the observed log ratios, and the genes are ranked by a single test statistic which incorporates simultaneous differential and equivalent gene expression testing.
\par
In Section \ref{sec:motivation}, our motivation for gene profiling is presented. Section \ref{sec:design} sets out the details of the experimental design for a pluripotent (stem cell) time  course experiment which provided our initial motivation for the ensuing methodological development. The theoretical underpinnings of gene profiling are described in Section \ref{sec:profile}, which entails a review of equivalence testing (Section \ref{sec:equiv}) and intersection-union tests (Section \ref{sec:IUT}). The gene profiling methodology is set out in Section \ref{sec:method}, and the results obtained from our application to a stem cell experiment are presented in Section \ref{sec:results}. In Section \ref{sec:discussion}, some further work and how to apply the methods in {\tt limma} are briefly discussed.


\section{Motivation: pluripotency}
\label{sec:motivation}
Our motivating example is a stem cell experiment originally conducted by the Rathjen laboratory, formerly of the University of Adelaide. The aim of the experiment was to identify genes associated with pluripotency in mice embryonic stem cells \citep{DAmour:2003,Ramalho-Santos:2002}. Early stem cells have the potential to differentiate into any body cell: a property known as pluripotency. This ability is present in mice stem cells up to and including day three. After this the stem cells become multipotent: they still have the ability to differentiate into different types of cells, but now a limited number. For example, haemopoetic stem cells can differentiate into blood cells but not nerve cells. As pluripotency is restricted to the early stem cells, day 3 or earlier, genes that have high expression levels in cells up to day 3, but low, or monotonically decreasing expression levels thereafter, are likely to be associated with the biochemical pathways involved in the pluripotency ability of these cells (personal communication, Dr Chris Wilkinson).


\subsection{Pluripotency example: experimental design}
\label{sec:design}
Stem cells were isolated from the early embryo and grown in culture dishes. The cells were allowed to replicate and grow over the medium in the dish. Once the cells had crowded the plate, they were removed, separated and plated onto new plates. This cycle of growth and re-plating is called a {\it passage}. The Rathjen laboratory isolated mice embryonic stem cells, and for this experiment, used cells from passages 21, 22, 23 and 24. The cells were stimulated to differentiate into multipotent cells, and on days 0, 3, 6 and 9 after stimulation, samples were taken and the messenger RNA (mRNA) obtained. 
\par
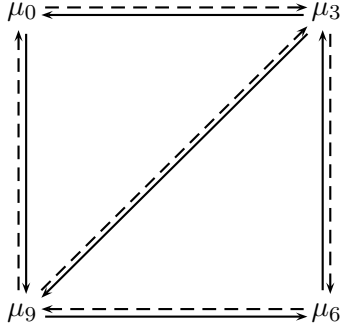
\begin{figure}[htbp]
\begin{center}
\psset{xunit=1cm,yunit=1cm}
\begin{pspicture} (0,0)(4,4)
\rput(0,4){$\mu_0$}
\rput(0,0){$\mu_9$}
\rput(4,4){$\mu_3$}
\rput(4,0){$\mu_6$}
\psline{->}(0.3,-0.05)(3.75,-0.05)
\psline[linestyle=dashed]{<-}(0.25,0.05)(3.7,0.05)
\psline[linestyle=dashed]{->}(0.3,4.05)(3.75,4.05)
\psline{<-}(0.25,3.95)(3.7,3.95)
\psline[linestyle=dashed]{->}(-0.05,0.3)(-0.05,3.75)
\psline{<-}(0.05,0.25)(0.05,3.7)
\psline{->}(3.95,0.3)(3.95,3.75)
\psline[linestyle=dashed]{<-}(4.05,0.25)(4.05,3.7)
\psline[linestyle=dashed]{->}(0.25,0.3)(3.75,3.8)
\psline{<-}(0.25,0.2)(3.75,3.7)
\end{pspicture}
\caption{Microarray comparisons made within each passage for the pluripotency stem cell experiment. Each arrow represents two arrays, one for each passage (passage 21/22 continuous arrow, passage 23/24 dashed arrow), with the arrow head pointing to the sample labeled with cy5, and the sample at the arrow tail labeled with cy3. Day 0, 3, 6, and 9 are represented by $\mu_0,\mu_3,\mu_6$, and $\mu_9$ respectively.}
\label{fig:design}
\end{center}
\end{figure}

The gene expressions of the 16 samples of stem cell mRNA for the four days (0, 3, 6, 9) and four passages, were measured. Within each passage, five comparisons were made, namely, day 0 to day 3, day 0 to day 9, day 3 to day 6, day 3 to day 9, and day 6 to day 9. The experimental design in terms of the true gene expression levels, $\boldsymbol{\mu}$ (see Section \ref{sec:profile}) is summarized in Figure \ref{fig:design}, while the experimental design in terms of the gene profiling parameters, $\boldsymbol{\gamma}$ (Section \ref{sec:profile}), is compared to the design in terms of the true expression levels in Table \ref{tab:parameter}. 
The clone library used in the experiment was the Compugen 22,000 mouse oligonucleotide library (http://www.microarray.adelaide.edu.au/libraries/). In total, 20 arrays were hybridized on two-colour long-oligonucleotide microarrays, with five arrays within each passage. The five arrays consisted of the five comparisons detailed above. In this analysis, the stem cells from each passage were treated as independent biological replicates.
\begin{table}[htbp]
	\begin{center}
		\begin{tabular}{ccc}
		\hline
		Day & Parameterization& Parameterization\\
		 & in terms of $\mu$'s & in terms of $\gamma$'s\\ 
		\hline
		0 & $\mu_0$ & $\gamma_0+\gamma_1+\gamma_2+\frac12\gamma_3$\\
		3 & $\mu_3$ & $\gamma_0+\gamma_1+\gamma_2-\frac12\gamma_3$\\
		6 & $\mu_6$ & $\gamma_0+\gamma_2$\\
		9 & $\mu_9$ & $\gamma_0$\\
		\hline
		\end{tabular}	
	\end{center}
	\caption{Parameterization of stem cell experiment in terms of absolute mean gene expressions ($\mu_i, i=0,3,6,9$) and in terms of the gene profile coefficients ($\gamma_i, i=0,1,2,3$).}
	\label{tab:parameter}
\end{table}


\section{Gene profiling methodology} 
\label{sec:profile}
\subsection{Development of method for stem cell experiment}
\label{sec:stem_cell}
\begin{figure}[!ht]
\begin{center}
\includegraphics{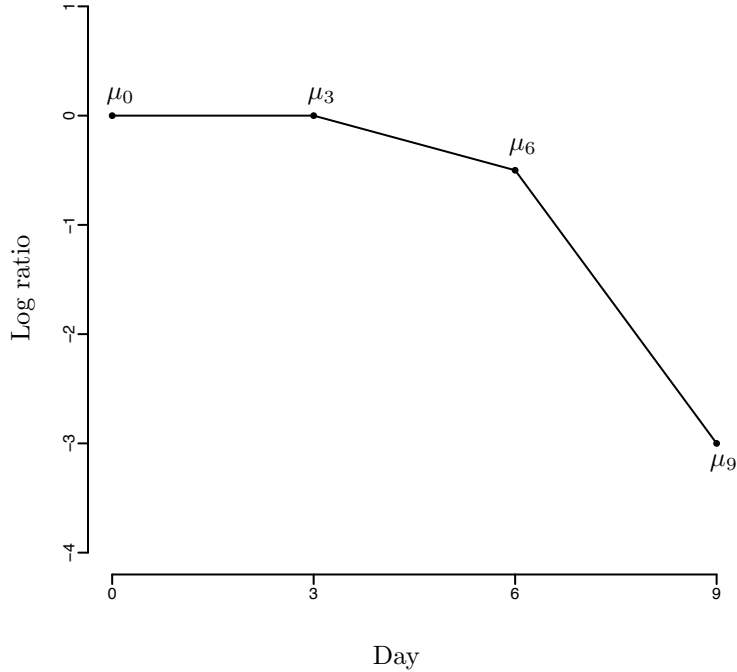}
\caption{The pre-specified gene expression profile for pluripotent genes. For each day, the log ratio with respect to day 0 is plotted.}
\label{fig:pluri}
\end{center}
\end{figure}

The expression criteria over time required for a pluripotent gene are:
\begin{itemize}
\item equal gene expression levels for days 0 and 3, 
\item higher gene expression levels for days 0 and 3 compared to day 9, and 
\item the gene expression level for day 6 to lie between the gene expression levels for day 0 and day 3, and the gene expression level for day 9.
\end{itemize}
The requisite (hypothetical) profile is illustrated in Figure \ref{fig:pluri}. 
\par
Consider the vector of true mean gene expression levels, $\boldsymbol{\mu}=\left(\mu_0,\mu_3,\mu_6,\mu_9\right)'$, where $\mu_i, i=0,3,6,9$ is the mean gene expression level on day $i$ as shown in Figure \ref{fig:design}. Since this is a vector in $\mathbb{R}^4$, it can be expressed as the linear combination of four linearly independent vectors. The first step in gene profiling is to choose vectors that represent the criteria for pluripotency. In the present example, this corresponds to
\begin{eqnarray}
\boldsymbol{\mu}=\gamma_0\left(\begin{array}{r}1 \\1 \\1 \\1\end{array}\right)+\gamma_1\left(\begin{array}{r}1 \\1 \\ 0 \\ 0\end{array}\right)+\gamma_2\left(\begin{array}{r}1 \\1 \\1 \\0\end{array}\right)+\gamma_3\left(\begin{array}{r}1/2 \\\mbox{-}1/2 \\0 \\0 \end{array}\right).\label{eq:mu_gamma}
\end{eqnarray}
With this choice of model, it follows that $\gamma_0=\mu_9, \gamma_1=(\mu_0+\mu_3)/2-\mu_6, \gamma_2=\mu_6-\mu_9,$ and $\gamma_3=\mu_0-\mu_3.$ Therefore, the pluripotent profile requires that $\gamma_1>0, \gamma_2>0, \gamma_3=0,$ but does not constrain $\gamma_0$. To find genes that achieve these criteria requires tests for equivalence as well as (simultaneous) tests for differential gene expression. In the next section, equivalence testing is discussed. We then describe how to simultaneously test for both differential and equivalent gene expression in a time course experiment.


\subsection{Statistical Equivalence}
\label{sec:equiv}

To determine pluripotency, it is necessary to demonstrate that $\gamma_3=0$. Conventional hypothesis testing is not applicable to this situation, but the equivalence testing  approach discussed in \cite{Wellek:2002} is.
\par
If $\boldsymbol{X}$ is a random vector whose probability distribution depends on a real-valued parameter $\theta$, then to test if $\theta$ is equivalent to zero, a neighbourhood around zero is constructed and the following null and alternative hypotheses are tested:
\begin{eqnarray}
H_0:&\left|\theta\right|\geq \epsilon,&\epsilon>0,\label{eq:equiv}\\
H_a:&\left|\theta\right|<\epsilon.&\nonumber
\end{eqnarray}
The neighbourhood defined by $\epsilon$ is the maximum that the parameter can vary and still be considered equivalent to zero. This neighbourhood is necessary to ensure that the power of the statistical test is greater than its significance level \citep{Wellek:2002}. 
\par
For the gene profiling model, the parameter $\epsilon$ is taken to be the largest that a gene's mean log ratio can vary around zero and not be of  ``significant'' gene expression, according to biologists. In practice, a working understanding of equivalent gene expression should be decided upon in advance in consultation with biologists. Unfortunately however, relatively little is known about gene-specific variation per se: information that could of course be used to decide on an appropriate value of $\epsilon$. We discuss potentially suitable choices of $\epsilon$ in Section \ref{sec:results}, but for the present we will assume an appropriate $\epsilon$ to be available. Using such a value of $\epsilon$, the simplest and most common way to test the hypotheses in (\ref{eq:equiv}) is via {\it Confidence Interval Inclusion} (CII). 
 \par
 Consider the null and alternative hypotheses specified in (\ref{eq:equiv}). We calculate a confidence interval, $R_\alpha(\boldsymbol{X})$, from the observed data $\boldsymbol{X}$, where
\begin{eqnarray}
R_\alpha(\boldsymbol{X})=\left(L_\alpha(\boldsymbol{X}),U_\alpha(\boldsymbol{X})\right); \label{eq:region}
\end{eqnarray}
$L_\alpha(\boldsymbol{X})$ and $U_\alpha(\boldsymbol{X})$ are random variables, such that
\begin{eqnarray*}
P\left(  \theta\in\left(  L_\alpha(\boldsymbol{X}),\infty\right)\right)=P\left(\theta\in\left(\mbox{-}\infty,U_\alpha(\boldsymbol{X})\right)\right)=1-\alpha.
\end{eqnarray*}
We reject the null hypothesis in favour of equivalence if and only if
\begin{eqnarray*}
R_\alpha(\boldsymbol{X})\subset(\mbox{-}\epsilon,\epsilon),
\end{eqnarray*}
i.e., the confidence interval is contained entirely within the interval $(\mbox{-}\epsilon,\epsilon)$. This is an $\alpha$-level test. 
\par
The equivalence formulation can be used to test that  $\gamma_3$ in equation (\ref{eq:mu_gamma}) is equivalent to zero with the following null and alternative hypotheses:
\begin{eqnarray}
H_0:\left|\gamma_3\right|\geq \epsilon &\mbox{ vs. } & H_a:\left|\gamma_3\right|< \epsilon.\label{eq:gamma}
\end{eqnarray}
For example, to test the hypotheses in (\ref{eq:gamma}), the confidence interval
$$
\left(\hat{\gamma}_3-t^*\mbox{SE}(\hat{\gamma}_3),\hat{\gamma}_3+t^*\mbox{SE}(\hat{\gamma}_3)\right),
$$ 
is calculated and $\gamma_3$ is concluded to be equivalent to zero if this confidence interval lies within $\left(\mbox{-}\epsilon,\epsilon\right)$. In this confidence interval, $t^*$ is chosen such that $P(T>t^*)=\alpha$, where $T$ has a $t$-distribution with the appropriate degrees of freedom for $\gamma_3$.  
\par
Confidence interval inclusion can also be used to (separately) test whether $\gamma_1$ and $\gamma_{2}$ are significantly positive. The null and alternative composite hypotheses for $\gamma_{1}$ are 
\begin{eqnarray*}
H_0:\gamma_1\leq 0 &\mbox{ vs. } & H_a:\gamma_1>0,
\end{eqnarray*}
and for $\gamma_{2}$ are
\begin{eqnarray*}
H_0:\gamma_2\leq 0 &\mbox{ vs. } & H_a:\gamma_2>0.
\end{eqnarray*}
\par
For an $\alpha$-level test here, a one-sided $(1-\alpha)100\%$ confidence interval for $\gamma_1$ is calculated:
\begin{eqnarray*}
\left(\hat{\gamma}_1-t^*\mbox{SE}(\hat{\gamma}_1),\infty\right),
\end{eqnarray*}
and if this interval is contained in $(0,\infty)$, $\gamma_1$ is concluded to be significantly positive. Similarly for $\gamma_{2}$.
\par
These methods allow testing of each criterion separately, but for pluripotency all three criteria need to be valid simultaneously. The authors' method to simultaneously test for both equivalence of parameters to zero and significant departures of parameters from zero is described in the next section.


\subsection{Intersection-Union test}
\label{sec:IUT}
The test for each criterion discussed in Section \ref{sec:equiv} can be incorporated simultaneously into a single null and a single alternative hypothesis as follows:
\begin{eqnarray}
H_0:&\left(\gamma_1\leq0\right) \;\bigcup\; \left(\gamma_2\leq0\right) \;\bigcup\; \left(\left|\gamma_3\right|\geq\epsilon\right),&\epsilon>0,\label{eq:null}\\
\mbox{versus } H_a:&\left(\gamma_1>0\right) \;\bigcap\; \left(\gamma_2>0\right) \;\bigcap\; \left(\left|\gamma_3\right|<\epsilon\right)\label{eq:alt}.&
\end{eqnarray}

\par
The hypotheses in (\ref{eq:null}) and (\ref{eq:alt}) represent an {\it intersection-union test} (IUT)\citep{Berger:1982}. To review, in an IUT, the null hypothesis is expressed as a union, 
$$
H_0:\theta\in\bigcup_{\gamma\in\Gamma}\Theta_\gamma,
$$ 
where $\Theta_\gamma$ is a subset of the parameter space indexed by $\gamma$. The rejection region $R$ of this IUT is of the form $R=\bigcap_{\gamma\in \Gamma}R_\gamma$, where $R_\gamma$ is the rejection region for a test of $H_{0 \gamma}:\theta\in\Theta_{\gamma}$ versus $H_{1 \gamma}:\theta\in\Theta_{\gamma}^{c}$. This is an $\alpha$-level test, where $\alpha=\mbox{sup}_{\gamma\in\Gamma}\alpha_\gamma$ and $\alpha_\gamma$ is the size of the test $H_{0 \gamma}$, with rejection region $R_\gamma$.
\par
Thus for each $\gamma_i, i=1,2,3$, in the null hypothesis statement (\ref{eq:null}), a test of size $\alpha_i$ is found, and the overall IUT will be of level $\mbox{sup } \alpha_i$. Using the confidence interval inclusion method discussed in the previous section to test each $\gamma_i$ separately, each test being of level $\alpha$, gives an overall $\alpha$-level test.
\par
Our main aim is to rank the genes in our motivating example according to their match with the pluripotent profile. The testing methodology described can be modified to give a quantitative measure of how closely each gene matches the desired profile. Considering each gene separately, for each parameter, $\gamma_i, i=1,2,3$, confidence interval inclusion is used to test the associated null hypothesis. Rather than using a fixed significance level, the smallest significance level, $\alpha_i$, for each $\gamma_i,i=1,2,3$ respectively, is found, such that the null hypothesis is rejected. The supremum of $\alpha_i, i=1,2,3$ is used as the test statistic to rank the genes. In fact, in the stem cell experiment, rather than calculate $\alpha_{i}$ for each $\gamma_{i}, i=1,2,3$, the width of the largest confidence interval, $U_{i}$, for each $\gamma_{i}$ that was contained within the rejection region was used. The infimum, $U,$ of the $U_{i}$ was then used to rank the genes (it should be noted that this is equivalent to ranking based on $\sup \alpha_i$). 
\par
\begin{figure}[htbp]
\begin{center}
\includegraphics[width=3in]{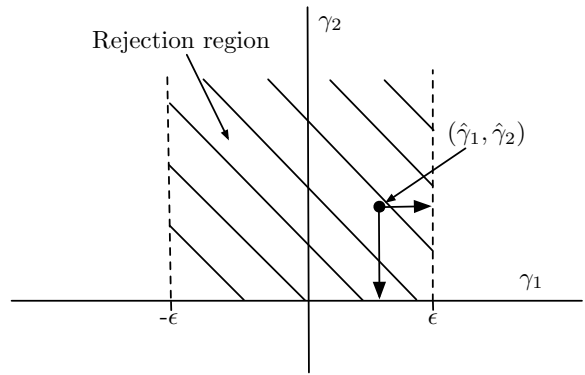}
\caption{Illustration: for each gene, the distance from $\left(\hat{\gamma}_1,\hat{\gamma}_2\right)$ to the nearest boundary of the rejection region is used to rank the genes.}
\label{fig:rej_region}
\end{center}
\end{figure}

 
To further elucidate the method, consider Figure \ref{fig:rej_region}. This illustrates a two-dimensional example where the criteria are $\gamma_1=0$ and $\gamma_2>0$. The rejection region is indicated by the rectangular shaded region. The point $\left(\hat{\gamma}_1,\hat{\gamma_2}\right)$ is the estimate of $\left(\gamma_1,\gamma_2\right)$. The distance to the nearest boundary of the rejection region is calculated in standard errors of the estimate and this distance is used to rank the genes, with larger values indicative of association with pluripotency. Genes whose profiles do not lie within the rejection region are excluded from the ranking.
\par
The above development leads to the general methodology for determining pluripotency described in the next section. 

\subsection{Gene profiling for pluripotency}
\label{sec:method}
The scanned images for each hybridized microarray slide were analysed using {\tt SPOT} \citep{Yang:2001} to give the cy3 and cy5 intensities for each gene \citep{Yang:2001,Adams:1994}. The data were then normalized by within-array print-tip loess, and the gene profile model was fitted to the normalized data using {\tt limma} \citep{Smyth:2005a} in {\tt R} \citep{R-Development-Core-Team:2006}. For each gene, the model parameter estimates and standard errors obtained by {\tt limma} were used to calculate the $U$ statistic (see below) using C code embedded in {\tt R} code. The genes were then ranked using the $U$ statistic. 
\par
The vector of observed log ratios $\boldsymbol{M}$ was expressed as a linear model of the true gene expression levels $\boldsymbol{\mu}$ as follows:
$$
\boldsymbol{M}=X^*\boldsymbol{\mu}+\boldsymbol{E},
$$
where $X^*$ is the design matrix representing the mRNA comparisons made on each array, and $\boldsymbol{E}$ is assumed to be distributed as $N_{20}(\boldsymbol{0},\sigma^2I)$. Using equation (\ref{eq:mu_gamma}) to substitute for $\boldsymbol{\mu}$, gives 
\begin{eqnarray*}
M&=&X^*
\left(\begin{array}{rrrr}1 & 1 & 1 & 1/2 \\1 & 1 & 1 & \mbox{-}1/2 \\1 & 0 & 1 & 0 \\1 & 0 & 0 & 0\end{array}\right)
\left(\begin{array}{c}\gamma_0 \\\gamma_1 \\\gamma_2 \\\gamma_3\end{array}\right)
+\boldsymbol{E}\\
&=&X\boldsymbol{\gamma}+\boldsymbol{E}.
\end{eqnarray*}
In the stem cell experiment, the microarray platform used was two-colour long oligonucleotide which, as for cDNA microarrays, measures relative gene expression, but not absolute gene expression levels. Therefore, the overall gene expression level, $\gamma_{0}$, could not be estimated and was removed from the model by changing the parameter vector to $(\gamma_1,\gamma_2,\gamma_3)$ and removing the first column of $X$.
\par
Estimates of $\boldsymbol{\gamma}$ were calculated via least squares, and the estimate of $\sigma^2$ was obtained using the empirical Bayes method utilized in {\tt limma}; this gives a robust posterior estimate of $\sigma^2$ based on a prior which ``borrows'' information from the observed variance of all the genes on the array.
\par
For each gene, three tests statistics, $U_1,U_2$ and $U_3$ were calculated as follows:
\begin{eqnarray*}
U_1=\frac{\hat{\gamma}_1}{\mbox{SE}(\hat{\gamma}_1)},\ 
U_2=\frac{\hat{\gamma}_2}{\mbox{SE}(\hat{\gamma}_2)},\ 
U_3=\frac{\epsilon-\left|\hat{\gamma}_3\right|}{\mbox{SE}(\hat{\gamma}_3)},
\end{eqnarray*}
where SE($\hat{\gamma}_i$) is the $i$th diagonal element of the square matrix: $s\sqrt{(X'X)^{\mbox{-}1}}$, and $s$ is the posterior estimate of $\sigma$. The minimum of $U_i, i=1,2,3$, is used to rank the genes, for which genes with larger values of $U$ are more likely to be associated with pluripotency.
\par
Genes whose estimate $(\hat{\gamma_1},\hat{\gamma_2},\hat{\gamma_3})$ of $(\gamma_1,\gamma_2,\gamma_3)$ did not lie within the rejection region, i.e. those genes for which at least one $U_i, i=1,2,3$ was negative, were excluded from the ranking.


\section{Application: determining genes associated with pluripotency using gene profiling}
\label{sec:results}
The model (\ref{eq:mu_gamma}) was fitted to the stem cell data with $\epsilon=1$. In addition, the test statistics were changed to test for $\gamma_2>1.5$, i.e., $U_2=\frac{\hat{\gamma}_2-1.5}{\mbox{SE}(\hat{\gamma}_2)}$. The value of $1.5$ was chosen to ensure a large difference between the gene expression levels on days 0, 3 and 6 compared with the gene expression level on day 9. 
\par
\begin{figure}[!ht]
\begin{center}
\includegraphics{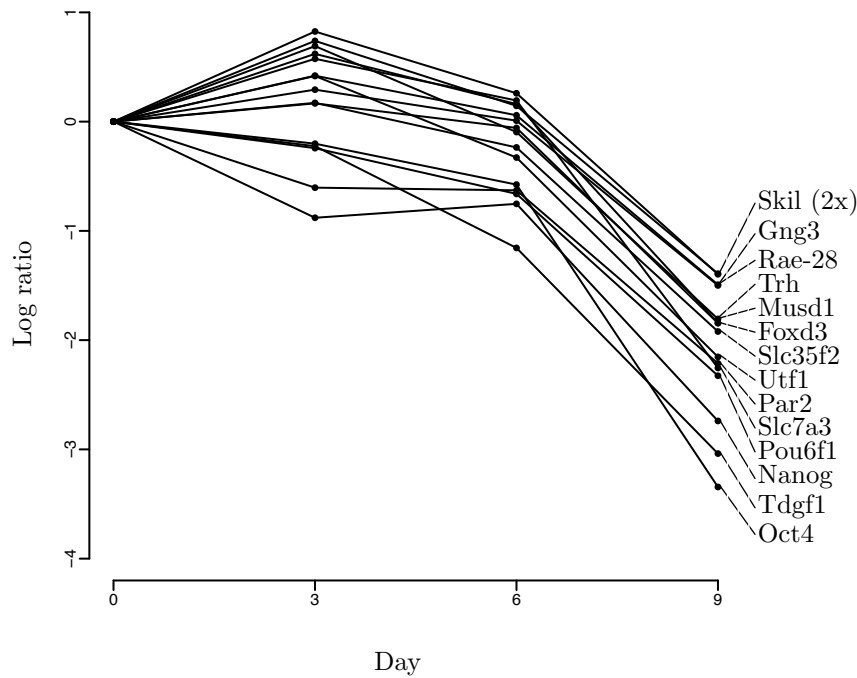}
\caption{Fitted log ratios with respect to day 0 for the ranked genes for the pluripotency profile (\ref{eq:mu_gamma}).}
\label{fig:pluri10}
\end{center}
\end{figure}

The ranked genes are given in Table \ref{tab:pluri}, and the fitted profiles for these 15 genes are shown in Figure \ref{fig:pluri10}. Figure \ref{fig:pluri10} shows the fitted log ratios with respect to day 0 for the four time points: day 0, day 3, day 6, and day 9. Therefore, all of the profiles will pass through zero on day 0. The profiles demonstrate the required trajectory: equal expression for day 0 and day 3, higher gene expression levels for days 0 and 3 compared to day 9, and the gene expression level for day 6 lying between the gene expression levels for days 0 and 3 and that for day 9.
\input{./Tables/Tab_1_pluri_anno.txt}
\par
The top-ranked gene, Oct4, is well-known to be associated with pluripotency \citep{Rodda:2005,Loh:2006} and would therefore be expected to appear amongst the top-ranked genes for pluripotency in this experiment. Other genes of note in the ranked genes in Table \ref{tab:pluri} are Utf1 (rank 2) which is associated with undifferentiated embryonic cell transcription \citep{Nishimoto:2005fj}, and Nanog (rank 11) which is central to embryonic stem cell pluripotency \citep{Wang:2006}.
\par
The recent article by \cite{Wang:2006} isolated proteins associated with the protein Nanog and thus with pluripotency. Of the 38 proteins discussed in \cite{Wang:2006}, Oct4 and Nanog appeared in our list of ranked genes using model (\ref{eq:mu_gamma}): ranks 1 and 11 respectively. The remaining proteins were not in the ranked genes as the profiles of the associated mRNAs are not consistent with profile (\ref{eq:mu_gamma}).
\par
\par

{\it Sensitivity analysis}: As stressed previously, the choice of the neighbourhood around zero assumed for equivalence (i.e., $\epsilon$) should be decided upon in consultation with biologists. However, this is problematic since biologists still have relatively little explicit knowledge of gene-wise expression variability, and therefore, what precisely and quantitatively may represent equivalence of gene expression.
\par
\begin{figure}[!ht]
\begin{center}
\includegraphics{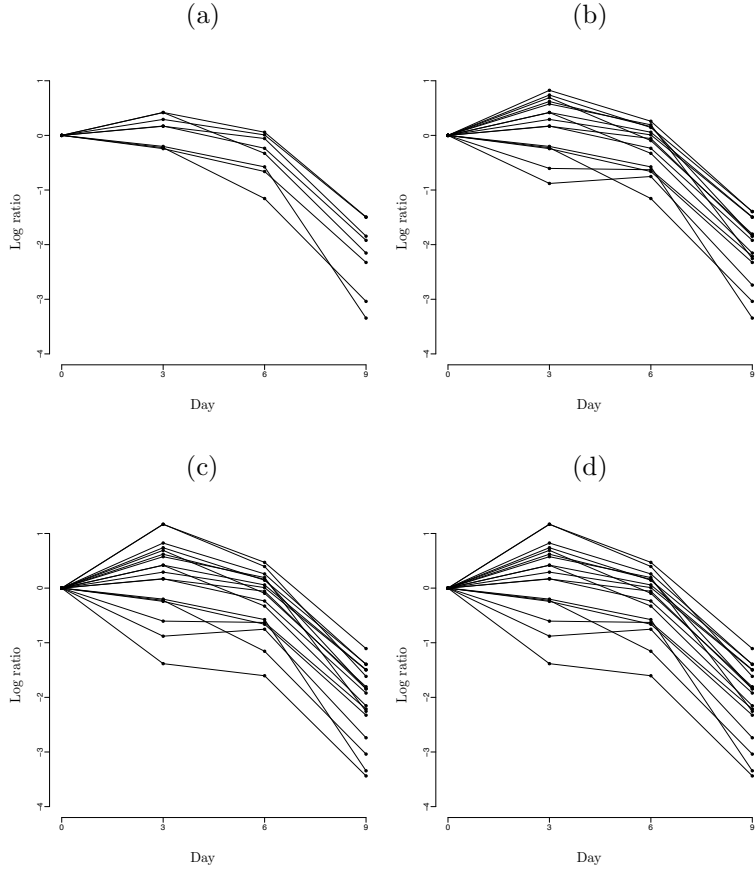}
\caption{Fitted log ratios with respect to day 0 for the ranked genes with (a) $\epsilon=$ 0.5, (b) $\epsilon=$ 1, (c) $\epsilon=$ 1.5, and (d) $\epsilon=$ 2.}
\label{fig:sensitivity}
\end{center}
\end{figure}

To investigate the potential effects of altering the neighbourhood defined by $\epsilon$, the primary analysis was repeated assuming, respectively, values of $\epsilon= 0.5,1,1.5, \mbox{ and }2.$ In Figure \ref{fig:sensitivity} the profiles for the genes which have observed profiles that lie within the rejection region are plotted for each choice of equivalence neighbourhood. As the equivalence neighbourhood width ($\epsilon$) increases, more genes have profiles that lie within the rejection region, but there is greater variation between the gene expression levels for day 0 and day 3. Nevertheless, gene profiling in this application has been demonstrated to be reasonably robust. For $\epsilon$=0.5, 1 and 1.5, Oct4 was ranked as the top gene, while for $\epsilon$=2, it had only dropped to rank 2. 
\par
\par

\begin{figure}[!ht]
\begin{center}
\includegraphics{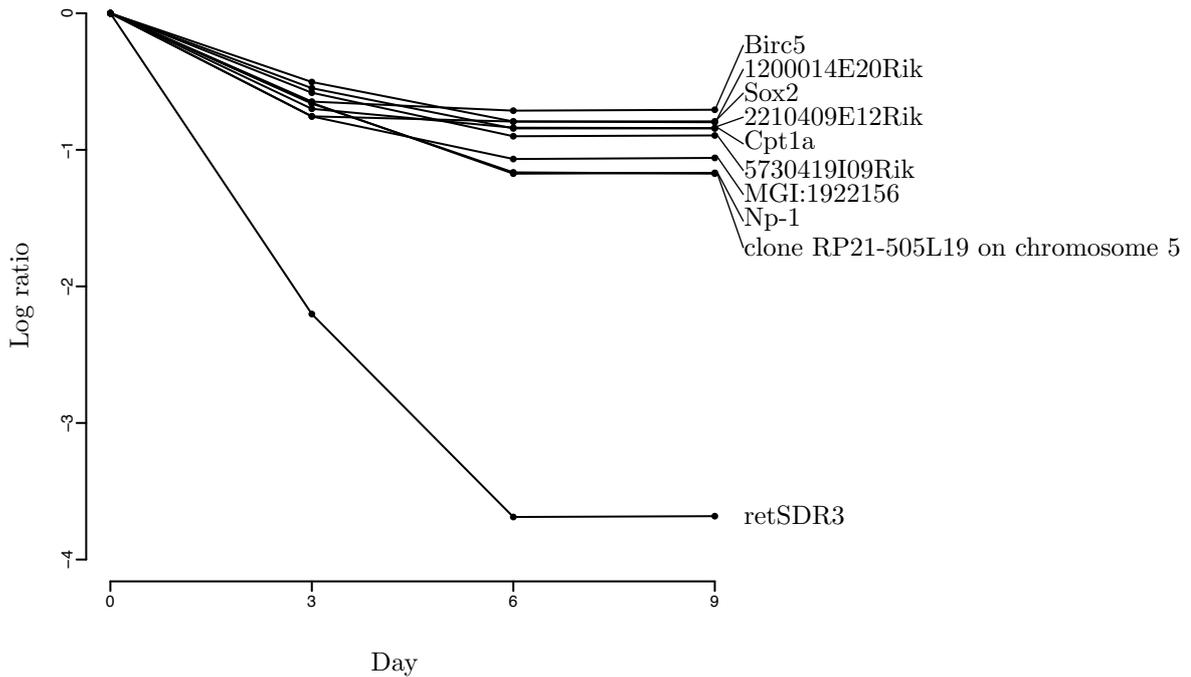}
\caption{Fitted log ratios with respect to day 0 for the top 10 ranked genes for the Sox2 profile.}
\label{fig:Sox2_results}
\end{center}
\end{figure}

{\it Profiling for Sox2}: It is well known that  the gene Sox2 is commonly associated with pluripotency \citep{Rodda:2005}, but it was not in the ranked genes using the gene expression profile (\ref{eq:mu_gamma}): the fitted gene expression profile for Sox2 is very different from the pluripotent profile used in the analysis. The criteria for the profile of Sox2 are:
higher gene expression level on day 0 compared to the gene expression levels for days 6 and 9;
equivalent gene expression levels on days 6 and 9; and 
the gene expression level for day 3 to lie between the gene expression level for day 0 and the levels for days 6 and 9. 
Gene profiling can be used to rank the genes according to these alternative criteria. An appropriate model for Sox2 is:
\begin{eqnarray*}
\boldsymbol{\mu}
=\left(\begin{array}{rrrr}1 & 1 & 1 & 0 \\1 & 0 & 1 & 0 \\1 & 0 & 0 & \frac12 \\1 & 0 & 0 & \mbox{-}\frac12\end{array}\right)
\boldsymbol{\gamma},
\end{eqnarray*}
in which 
$\gamma_0$ is unrestrained,
$\gamma_1>0$,
$\gamma_2>0$, and
$\gamma_3$ is equivalent to zero. This model was fitted to the data and the ranked genes are shown in Figure \ref{fig:Sox2_results}. The ranked genes were Cpt1a, 1200014E20Rik, 2210409E12Rik, Sox-2, Np-1, Birc5, 5730419I09Rik, MGI:1922156, retSDR3, and clone RP21-505L19 on chromosome 5. Sox2 was ranked at position 4. Of note is retSDR3. This gene has the required form with a larger difference in gene expression between day 0 and days 6 and 9 compared to the other genes. Even with this large difference, retSDR3 is low down in the ranking at rank 9. This low ranking is because retSDR3 has a large gene expression variance (0.181) compared to the other ranked genes (average gene expression variance of 0.054). This illustrates that if two genes have the same coefficient values, gene profiling will rank lower the gene which has the larger variance and thus more uncertainty about its true profile.


\section{Discussion of further work}
\label{sec:discussion}
In general, gene profiles of interest to molecular biologists often consist of two types of criteria: equal gene expression at different time points and differential gene expression at different time points. Gene profiling provides a straightforward methodology to filter genes which satisfy these two types of criteria simultaneously. We believe that this has not been accomplished using previously available techniques. By simultaneously testing for all criteria, gene profiling effectively filters out and excludes genes that are only partially consistent with the required profile. We now touch on some areas requiring further work.
\par
{\it Choice of $\epsilon$}: As noted in Section \ref{sec:IUT}, to test for a parameter being equal to zero, a neighbourhood of width $\epsilon$ is defined. This neighbourhood is the amount that the parameter could vary and still be considered equivalent to zero. In this paper, the choice of $\epsilon$ was based on plotting profiles for the various choices of $\epsilon$, and choosing the best $\epsilon$ to give the required pre-specified profiles. Ideally, the choice of $\epsilon$ should be based on consultation with biologists, to the extent that such knowledge is available. One would anticipate that such requisite knowledge will gradually accrue over time, as microarray and other new genomics technologies are more widely applied in molecular biology and genetics.
\par
\begin{figure}[htbp]
\begin{center}
\includegraphics{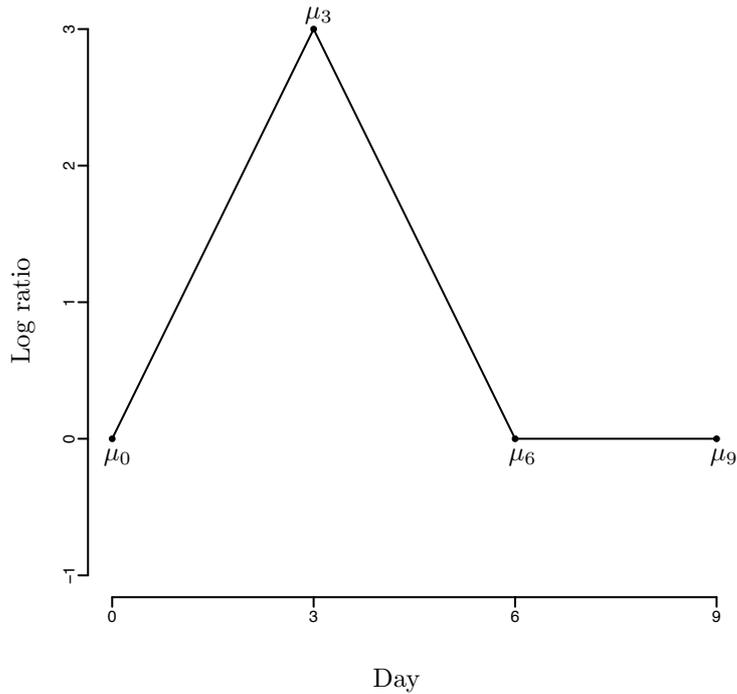}
\caption{Hypothetical gene expression profile for a day 3 marker gene.}
\label{fig:marker3}
\end{center}
\end{figure}

{\it Invariance of parameterization}: Another area requiring further research is the invariance (or otherwise) of reparameterization. In his (2002) book, Wellek notes:
\begin{quotation}
``... in contrast to the corresponding conventional testing problems with the common boundary of null and alternative hypothesis [{\it sic}] being given by zero, equivalence problems remain generally not invariant under redefinitions of the main parameter.''
\end{quotation}
To illustrate this point, consider the problem of finding marker genes for day 3 in the stem cell experiment. The criteria for such genes are:
high gene expression level on day 3, as well as
equal and low gene expression levels on day 0, day 6 and day 9.
The requisite profile is illustrated in Figure \ref{fig:marker3}. Examination of the profile reveals three possible models:
$$
\boldsymbol{\mu}=\left(\begin{array}{rrrr}1 & 0 & 0 & 0 \\1 & 1 & \mbox{-}\frac13 & \mbox{-}\frac13 \\1 & 0 & 0 & \mbox{-}1 \\1 & 0 & \mbox{-}1 & 0\end{array}\right)\boldsymbol{\gamma},
\boldsymbol{\mu}=\left(\begin{array}{rrrr}1 & 0 & 0 & 0 \\1 & 1 & \mbox{-}\frac23 & \frac13 \\1 & 0 & \mbox{-}1 & 1 \\1 & 0 & \mbox{-}1 & 0\end{array}\right)\boldsymbol{\gamma},
\boldsymbol{\mu}=\left(\begin{array}{cccc}1 & 0 & 0 & 0 \\1 & 1 & \mbox{-}\frac23 & \mbox{-}\frac13 \\1 & 0 & \mbox{-}1 & 0 \\1 & 0 & \mbox{-}1 & \mbox{-}1\end{array}\right)\boldsymbol{\gamma},
$$
where $\boldsymbol{\gamma}=(\gamma_0,\gamma_1,\gamma_2,\gamma_3)'$
with
$\gamma_0$ unrestrained,
$\gamma_1$ significantly positive,
$\gamma_2$ equivalent to zero, and
$\gamma_3$ equivalent to zero.
\par
The three models may not necessarily give the same results. This is because equivalence is not transitive, i.e., if $\mu_{0}$ is equivalent to $\mu_{6}$, and $\mu_{6}$ is equivalent to $\mu_{9}$, it is not necessarily true that $\mu_{0}$ is equivalent to $\mu_{9}$. This is because equivalence is defined in a neighbourhood and so a ``drift'' resulting in $\mu_{0}$ and $\mu_{9}$ being too far apart to be considered equivalent can occur. Methods to impose invariance are currently under investigation by the authors. Although this is an interesting area of research, invariance of reparameterization does not limit the use of gene profiling. There are many pre-existing statistical tests, e.g., the Wald test, that are not invariant under reparameterization. In many cases, the research hypotheses will dictate the optimal model to use.
\par
{\it Applying gene profiling using {\tt limma}}: Gene profiling is easily implemented by fitting the model to the data using {\tt limma} and then calculating the $U$ statistics. The calculation of the $U$ statistics was written in C to decrease the run time, but is easy to implement in {\tt R}. 
\par
To conclude, gene profiling introduces a flexible method to select genes for a pre-specified time-course profile. Gene profiling is straightforward to implement in practice, requiring only small modifications to the {\tt R} package {\tt limma}, and can be used to select for most profiles of interest to biologists. The application of gene profiling in this article has been to two-colour microarrays, but it could readily be modified for use for other microarrays platforms, such as Affymetrix GeneChip \citep{Lockhart:1996}, and for other technologies where it is required to rank observations by correspondence with a pre-specified profile.

\section{Acknowledgements}
We thank the Rathjen group for the use of the stem cell data, and Dr Chris Wilkinson and Professor Terry Speed for useful information about stem cell experiments.  The second- and third-named authors are grateful to the Australian Research Council for research support through a Discovery Project Grant.  The first-named author was supported by a George Fraser PhD scholarship at the University of Adelaide.

\newpage
\listoffigures
\newpage
\listoftables

\end{document}

%% file: Tables/Tab_1_pluri_anno.txt
\begin{table}[!ht]
\begin{center}
\begin{tabular}{lrrrr}
  \hline
Gene Names & $\hat{\gamma}_1$ & $\hat{\gamma}_2$ & $\hat{\gamma}_3$ & U \\
  \hline
Oct4		& 0.48 & 2.77 & 0.20 & 2.53 \\
Utf1 		& 0.54 & 1.82 & $-$0.42 & 2.46 \\
Tdgf1 		& 1.04 & 1.88 & 0.22 & 1.80 \\
Slc35f2 	& 0.32 & 1.69 & $-$0.17 & 1.53 \\
Trh		 	& 0.44 & 1.71 & $-$0.69 & 1.50 \\
Foxd3 	& 0.14 & 1.79 & $-$0.17 & 1.33 \\
Musd1  	& 0.15 & 2.00 & $-$0.62 & 1.17 \\
Skil   	& 0.15 & 1.66 & $-$0.83 & 1.16 \\
Pou6f1 	& 0.54 & 1.66 & 0.24 & 1.13 \\
Par2   	& 0.33 & 1.58 & 0.60 & 0.75 \\
Nanog	& 0.31 & 1.99 & 0.88 & 0.69 \\
Slc7a3 	& 0.09 & 2.45 & $-$0.58 & 0.67 \\
Gng3 	& 0.15 & 1.55 & $-$0.42 & 0.33 \\
Skil 	& 0.23 & 1.54 & $-$0.74 & 0.28 \\
Rae-28 	& 0.14 & 1.51 & $-$0.29 & 0.08 \\
   \hline
\end{tabular}
\caption{The ranked genes from fitting pluripotent profile (\ref{eq:mu_gamma}) to the stem cell data.}
\label{tab:pluri}
\end{center}
\end{table}

%% file: 11Geneprofilingpaper.bbl
\begin{thebibliography}{32}
\providecommand{\natexlab}[1]{#1}
\expandafter\ifx\csname urlstyle\endcsname\relax
  \providecommand{\doi}[1]{doi:\discretionary{}{}{}#1}\else
  \providecommand{\doi}{doi:\discretionary{}{}{}\begingroup
  \urlstyle{rm}\Url}\fi

\bibitem[{Adams and Bischof(1994)}]{Adams:1994}
\textsc{Adams, R. and Bischof, L.} (1994).
\newblock Seeded region growing.
\newblock \emph{IEEE Transactions on Pattern Analysis and Machine Intelligence}
  \textbf{16}, 641--647.

\bibitem[{Ahnert \emph{et~al.}(2006)Ahnert, Willbrand, Brown and
  Fink}]{Ahnert:2006}
\textsc{Ahnert, S.~E., Willbrand, K., Brown, F. C.~S. and Fink, T. M.~A.}
  (2006).
\newblock Unbiased pattern detection in microarray data series.
\newblock \emph{Bioinformatics} \textbf{22}, 1471--1476.

\bibitem[{Bar-Joseph \emph{et~al.}(2003)Bar-Joseph, Gerber, Gifford, Jaakkola
  and Simon}]{Bar-Joseph:2003}
\textsc{Bar-Joseph, Z., Gerber, G.~K., Gifford, D.~K., Jaakkola, T.~S. and
  Simon, I.} (2003).
\newblock Continuous representations of time-series gene expression data.
\newblock \emph{J Comput Biol} \textbf{10}, 341--56.

\bibitem[{Berger(1982)}]{Berger:1982}
\textsc{Berger, R.~L.} (1982).
\newblock Multiparameter hypothesis testing and acceptance sampling.
\newblock \emph{Technometrics} \textbf{24}, 295--300.

\bibitem[{Brown \emph{et~al.}(2006)Brown, Wilkinson, Waterman, Kok, Salerno,
  Diakiw, Reynolds, Scott, Tsykin, Glonek, Goodall, Solomon, Gonda and
  D'Andrea}]{Brown:2006}
\textsc{Brown, A.~L., Wilkinson, C.~R., Waterman, S.~R., Kok, C.~H., Salerno,
  D.~G., Diakiw, S.~M., Reynolds, B., Scott, H.~S., Tsykin, A., Glonek, G.~F.,
  Goodall, G.~J., Solomon, P.~J., Gonda, T.~J. and D'Andrea, R.~J.} (2006).
\newblock Genetic regulators of myelopoiesis and leukemic signaling identified
  by gene profiling and linear modeling.
\newblock \emph{Journal of Leukocyte Biology} \textbf{80}, 1--15.

\bibitem[{Celeux \emph{et~al.}(2005)Celeux, Martin and Lavergne}]{Celeux:2005}
\textsc{Celeux, G., Martin, O. and Lavergne, C.} (2005).
\newblock Mixture of linear mixed models for clustering gene expression
  profiles from repeated microarray experiments.
\newblock \emph{Statistical Modelling} \textbf{5}, 243--267.

\bibitem[{D'Amour and Gage(2003)}]{DAmour:2003}
\textsc{D'Amour, K.~A. and Gage, F.~H.} (2003).
\newblock Genetic and functional differences between multipotent neural and
  pluripotent embryonic stem cells.
\newblock \emph{PNAS} \textbf{100}, 11866--11872.

\bibitem[{Dudoit \emph{et~al.}(2002)Dudoit, Yang, Callow and
  Speed}]{Dudoit:2002}
\textsc{Dudoit, S., Yang, Y.~H., Callow, M.~J. and Speed, T.~P.} (2002).
\newblock Statistical methods for identifying differentially expressed genes in
  replicated c{DNA} microarray experiments.
\newblock \emph{Statistica Sinica} \textbf{12}, 111--139.

\bibitem[{Eisen \emph{et~al.}(1998)Eisen, Spellman, Brown and
  Botstein}]{Eisen:1998}
\textsc{Eisen, M.~B., Spellman, P.~T., Brown, P.~O. and Botstein, D.} (1998).
\newblock Cluster analysis and display of genome-wide expression patterns.
\newblock \emph{PNAS} \textbf{95}, 14863--14868.

\bibitem[{Ernst \emph{et~al.}(2005)Ernst, Nau and Bar-Joseph}]{Ernst:2005}
\textsc{Ernst, J., Nau, G.~J. and Bar-Joseph, Z.} (2005).
\newblock Clustering short time series gene expression data.
\newblock \emph{Bioinformatics} \textbf{21}, 159--168.

\bibitem[{Fleury \emph{et~al.}(2002)Fleury, Hero, Yoshida, Carter, Barlow and
  Swaroop}]{Fleury:2002}
\textsc{Fleury, G., Hero, A., Yoshida, S., Carter, T., Barlow, C. and Swaroop,
  A.} (2002).
\newblock Pareto analysis for gene filtering in microarray experiments.
\newblock \emph{Proc. XI European Signal Processing Conference} .

\bibitem[{Ghosh and Chinnaiyan(2002)}]{Ghosh:2002}
\textsc{Ghosh, D. and Chinnaiyan, A.~M.} (2002).
\newblock Mixture modelling of gene expression data from microarray
  experiments.
\newblock \emph{Bioinformatics} \textbf{18}, 275--286.

\bibitem[{Glonek and Solomon(2004)}]{Glonek:2004}
\textsc{Glonek, G.~F. and Solomon, P.~J.} (2004).
\newblock Factorial and time course designs for c{DNA} microarray experiments.
\newblock \emph{Biostatistics} \textbf{5}, 89--111.

\bibitem[{Hero and Fleury(2004)}]{Hero:2004}
\textsc{Hero, A.~O. and Fleury, G.} (2004).
\newblock Pareto-optimal methods for gene ranking.
\newblock \emph{The Journal of VLSI Signal Processing} \textbf{38}, 259--275.

\bibitem[{Lockhart \emph{et~al.}(1996)Lockhart, Dong, Byrne, Follettie, Gallo,
  Chee, Mittmann, Wang, Kobayashi, Horton and Brown}]{Lockhart:1996}
\textsc{Lockhart, D.~J., Dong, H., Byrne, M.~C., Follettie, M.~T., Gallo,
  M.~V., Chee, M.~S., Mittmann, M., Wang, C., Kobayashi, M., Horton, H. and
  Brown, E.~L.} (1996).
\newblock Expression monitoring by hybridization to high-density
  oligonucleotide arrays.
\newblock \emph{Nat Biotechnol} \textbf{14}, 1675--80.

\bibitem[{Loh \emph{et~al.}(2006)Loh, Wu, Chew, Vega, Zhang, Chen, Bourque,
  George, Leong, Liu, Wong, Sung, Lee, Zhao, Chiu, Lipovich, Kuznetsov, Robson,
  Stanton, Wei, Ruan, Lim and Ng}]{Loh:2006}
\textsc{Loh, Y.-H., Wu, Q., Chew, J.-L., Vega, V.~B., Zhang, W., Chen, X.,
  Bourque, G., George, J., Leong, B., Liu, J., Wong, K.-Y., Sung, K.~W., Lee,
  C. W.~H., Zhao, X.-D., Chiu, K.-P., Lipovich, L., Kuznetsov, V.~A., Robson,
  P., Stanton, L.~W., Wei, C.-L., Ruan, Y., Lim, B. and Ng, H.-H.} (2006).
\newblock The {O}ct4 and {N}anog transcription network regulates pluripotency
  in mouse embryonic stem cells.
\newblock \emph{Nat Genet} \textbf{38}, 431--40.

\bibitem[{L\"onnstedt \emph{et~al.}(2003)L\"onnstedt, Grant, Begley and
  Speed}]{Lonnstedt:2006}
\textsc{L\"onnstedt, I., Grant, S., Begley, G. and Speed, T.} (2003).
\newblock Microarray analysis of two interacting treatments: a linear model and
  trends in expression over time.
\newblock In Ingrid L\"onnstedt's Ph.D. thesis.

\bibitem[{McLachlan \emph{et~al.}(2006)McLachlan, Bean and
  Jones}]{McLachlan:2006}
\textsc{McLachlan, G.~J., Bean, R.~W. and Jones, L. B.~T.} (2006).
\newblock A simple implementation of a normal mixture approach to differential
  gene expression in multiclass microarrays.
\newblock \emph{Bioinformatics} \textbf{22}, 1608--1615.

\bibitem[{Nguyen \emph{et~al.}(2002)Nguyen, Arpat, Wang and
  Carroll}]{Nguyen:2002}
\textsc{Nguyen, D.~V., Arpat, A.~B., Wang, N.~Y. and Carroll, R.~J.} (2002).
\newblock {DNA} microarray experiments: Biological and technological aspects.
\newblock \emph{Biometrics} \textbf{58}, 701--717.

\bibitem[{Nishimoto \emph{et~al.}(2005)Nishimoto, Miyagi, Yamagishi, Sakaguchi,
  Niwa, Muramatsu and Okuda}]{Nishimoto:2005fj}
\textsc{Nishimoto, M., Miyagi, S., Yamagishi, T., Sakaguchi, T., Niwa, H.,
  Muramatsu, M. and Okuda, A.} (2005).
\newblock Oct-3/4 maintains the proliferative embryonic stem cell state via
  specific binding to a variant octamer sequence in the regulatory region of
  the {UTF}1 locus.
\newblock \emph{Molecular and Cellular Biology} \textbf{25}, 5084--5094.

\bibitem[{{R Development Core Team}(2006)}]{R-Development-Core-Team:2006}
\textsc{{R Development Core Team}} (2006).
\newblock \emph{R: A Language and Environment for Statistical Computing}.
\newblock R Foundation for Statistical Computing, Vienna, Austria.

\bibitem[{Ramalho-Santos \emph{et~al.}(2002)Ramalho-Santos, Yoon, Matsuzaki,
  Mulligan and Melton}]{Ramalho-Santos:2002}
\textsc{Ramalho-Santos, M., Yoon, S., Matsuzaki, Y., Mulligan, R.~C. and
  Melton, D.~A.} (2002).
\newblock ``{S}temness'': Transcriptional profiling of embryonic and adult stem
  cells.
\newblock \emph{Science} \textbf{298}, 597--600.

\bibitem[{Rodda \emph{et~al.}(2005)Rodda, Chew, Lim, Loh, Wang, Ng and
  Robson}]{Rodda:2005}
\textsc{Rodda, D.~J., Chew, J.-L., Lim, L.-H., Loh, Y.-H., Wang, B., Ng, H.-H.
  and Robson, P.} (2005).
\newblock Transcriptional regulation of {N}anog by {O}ct4 and {S}ox2.
\newblock \emph{J Biol Chem} \textbf{280}, 24731--7.

\bibitem[{Smyth \emph{et~al.}(2003)Smyth, Yang and Speed}]{Smyth:2003}
\textsc{Smyth, G., Yang, Y. and Speed, T.} (2003).
\newblock Statistical issues in c{DNA} microarray data analysis.
\newblock \emph{Methods Mol Biol} \textbf{224}, 111--36.

\bibitem[{Smyth(2005)}]{Smyth:2005a}
\textsc{Smyth, G.~K.} (2005).
\newblock Limma: linear models for microarray data pp. 397--420.

\bibitem[{Tai and Speed(2005)}]{Tai:2005}
\textsc{Tai, C.~Y. and Speed, T.~P.} (2005).
\newblock A multivariate empirical {B}ayes statistic for replicated microarray
  time course data.
\newblock Technical Report 667.

\bibitem[{Tamayo \emph{et~al.}(1999)Tamayo, Slonim, Mesirov, Zhu, Kitareewan,
  Dmitrovsky, Lander and Golub}]{Tamayo:1999}
\textsc{Tamayo, P., Slonim, D., Mesirov, J., Zhu, Q., Kitareewan, S.,
  Dmitrovsky, E., Lander, E.~S. and Golub, T.~R.} (1999).
\newblock Interpreting patterns of gene expression with self-organizing maps:
  Methods and application to hematopoietic differentiation.
\newblock \emph{PNAS} \textbf{96}, 2907--2912.

\bibitem[{Tavazoie \emph{et~al.}(1999)Tavazoie, Hughes, Campbell, Cho and
  Church}]{Tavazoie:1999}
\textsc{Tavazoie, S., Hughes, J.~D., Campbell, M.~J., Cho, R.~J. and Church,
  G.~M.} (1999).
\newblock Systematic determination of genetic network architecture.
\newblock \emph{Nat Genet} \textbf{22}, 281--285.

\bibitem[{Wang \emph{et~al.}(2006)Wang, Rao, Chu, Shen, Levasseur, Theunissen
  and Orkin}]{Wang:2006}
\textsc{Wang, J., Rao, S., Chu, J., Shen, X., Levasseur, D.~N., Theunissen,
  T.~W. and Orkin, S.~H.} (2006).
\newblock A protein interaction network for pluripotency of embryonic stem
  cells.
\newblock \emph{Nature} \textbf{444}, 364--368.

\bibitem[{Wellek(2002)}]{Wellek:2002}
\textsc{Wellek, S.} (2002).
\newblock \emph{Testing Statistical Hypotheses of Equivalence}.
\newblock CRC Press.

\bibitem[{Yang \emph{et~al.}(2001)Yang, Buckley and Speed}]{Yang:2001}
\textsc{Yang, Y., Buckley, M. and Speed, T.} (2001).
\newblock Analysis of c{DNA} microarray images.
\newblock \emph{Brief Bioinform} \textbf{2}, 341--9.

\bibitem[{Yeung \emph{et~al.}(2001)Yeung, Fraley, Murua, Raftery and
  Ruzzo}]{Yeung:2001}
\textsc{Yeung, K.~Y., Fraley, C., Murua, A., Raftery, A.~E. and Ruzzo, W.~L.}
  (2001).
\newblock Model-based clustering and data transformations for gene expression
  data.
\newblock \emph{Bioinformatics} \textbf{17}, 977--987.

\end{thebibliography}
